\documentclass[aps,prl,reprint,twocolumn,floatfix]{revtex4-1}
\usepackage{stmaryrd}
\usepackage{amsmath}
\usepackage{txfonts}
\usepackage{amssymb}
\usepackage{graphicx}
\usepackage{bm}
\usepackage{color}

\begin{document}
\title{Novel Topological Phase with Zero Berry Curvature}
\author{Feng Liu$^1$}
\author{Katsunori Wakabayashi$^{1,2}$}
\affiliation{$^1$Department of Nanotechnology for Sustainable Energy, School of Science and Technology,
Kwansei Gakuin University, Gakuen 2-1, Sanda 669-1337, Japan\\
$^2$National Institute for Materials Science (NIMS), Namiki 1-1, Tsukuba 305-0044, Japan}

\begin{abstract}
We present a two-dimensional (2D) lattice model that exhibits a nontrivial topological phase in the absence of the Berry curvature. Instead, the Berry connection provides the topological nontrivial  phase in the model, whose integration over the momentum space, the so called 2D Zak phase, yields a fractional wave polarization in each direction. These fractional wave polarizations manifest themselves as degenerated edge states with opposite parities in the model.
\end{abstract}

\maketitle

Topological insulators (TIs) hold robust edge states when they are put next to vacuum, which render them  promising for applications in electronics and quantum computing \cite{Hasan2010, Qi2011, Bansil2016}.
So far, three types of TIs are proposed: Chern insulators \cite{Klitzing1980, Haldane1988},  time-reversal-invariant (TRI) TIs \cite{Kane2005, Bernevig2006}, and topological crystalline insulators \cite{Fu2011}.
In these systems, the Bloch states generate nonvanishing Berry curvatures (BC), leading to the nontrivial topological indices after momentum integration of the BC \cite{Qi2008, Haldane2006, Shiozaki2014, Dong2016}.
The BC is the geometric analogue of the magnetic field in the momentum space \cite{Xiao2010}.
However, the nonzero BC is not always the necessary condition for the nontrivial topology of energy bands, if one recalls the situation of Aharonov-Bohm (AB) effect \cite{book2}, or a more relevant case of so-called Molecular Aharnovo-Bohm (MAB) effect \cite{Mead1979}, where the electrons are affected by the vector potential in spite of the zero magnetic field (in the AB effect) or zero Berry curvature (in the MAB effect) \cite{Mead1992}.
The geometric analogue of the vector potential in momentum space corresponds to the Berry connection.
Thus, the following question naturally raises up, can a similar situation happen in the momentum space that the Berry connection leads to nontrivial topology of energy bands in the absence of BC?

In this Letter, we show nontrivial topological phases in a 2D lattice system in the absence of BC
on the basis of Su-Schrieffer-Heeger (SSH) model \cite{Hughes2011} with Peierls distortions \cite{Lau2015, Lau2016}.
Here, we characterize the nontrivial topological phases by the integration of Berry connection over the first Brillouin zone (BZ), which is the 2D Zak phase accompanying with the fractional wave polarization in each direction \cite{Zak1989}. Different from the TIs such as Chern insulators, TRI TIs and topological crystalline insulators, Berry curvature in this system vanishes everywhere in BZ due to the coexistence of TR and inversion symmetries, except at locations of energy degeneracy, and furthermore there are no spin or spin-like degree of freedom in this system. Nontrivial topological edge states associated with this 2D Zak phases are manifested.
\begin{figure}[h]
\begin{center}
\leavevmode
\includegraphics[clip=true,width=0.75\columnwidth]{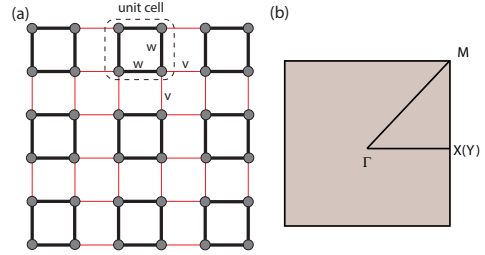}
\caption{(a) Schematic of 2D SSH model on a square lattice with symmetric hopping in each direction. Red and black bonds represent the intracelluar and intercelluar hoppings, respectively. The amplitudes of intracelluar and intercelluar hoppings are w and v, respectively. There are four atoms in one unit cell. (b) First BZ of the square lattice. }
\end{center}
\end{figure}

The Hamiltonian we consider is given as follows
\begin{equation}
\hat{\mathcal{H}}=\sum_{i,j}\left[(t_x+\delta t_x x_{i,j})c^\dagger_{i+1,j}c_{i,j}+
							 (t_y+\delta t_y y_{i,j})c^\dagger_{i,j+1}c_{i,j}\right]+h.c.,
\end{equation}
where $(i,j)$ represents a lattice point in a square lattice, $c^\dagger$ and $c$ are the creation and annihilation operators of a spinless electron at the site $(i,j)$, $t_x$ and $t_y$ are the transfer integral of the equidistant lattice (without distortions) along x- and y-directions, respectively, $\delta t_x$ and $\delta t_y$ are the electron-lattice coupling constants. The Peierls distortions are expressed in the following form
\begin{equation}
x_{i,j}=(-1)^{i}, \quad y_{i,j}=(-1)^j.
\end{equation}
According to the parity of site index $(i, j)$, there are two types of hopping in $x$ ($y$) direction such as $t_x-\delta t_x$ ($t_y-\delta t_y$) and $t_x+\delta t_x$ ($t_y+\delta t_y$). One can regard the two types of hopping in each direction as intracellular and intercellular hopping, respectively. Thus there are four atoms in one unit cell as displayed in Fig.~1(a), where the black and red bonds represent the intracellular and intercellular hopping, respectively. For simplicity, we set $t_x=t_y$, $\delta t_x=\delta t_y$ and define $w=t-\delta t$, $v=t+\delta t$. It is evident that our system respects both TR and inversion symmetries.

Applying Fourier transformation for the four atoms in Eq.~(1), one obtains a $4\times 4$ matrix $H_{ij}(k)$ in the momentum space, where $H_{12}=H_{34}=w+v\exp{\left(-ik_x\right)}$, $H_{13}=H_{24}=w+v\exp{\left(-ik_y\right)}$, and others are zeros and conjugated components. It is noticed that this Hamiltonian has a zero trace, which is due to the sublattice symmetry. The energy spectrum for $(w,v)=(2.0,1.0)$ [$(w,v)=(1.0, 2.0)$] is displayed in Fig. 2(a). There are four energy bands, two of them are doubly degenerate at $C_{4v}$ invariant points, and two of them are isolated.
Due to the continuous rotational symmetry, each energy band carries an integer angular momentum $j$ (as we consider spinless electrons) and its parity $\eta$ associated with rotation $\pi$ is given by $(-1)^j$. Because we consider square lattice with symmetric hopping, the system also respects $C_{4v}$ point group symmetry (PGS). Thus energy bands with negative parity are doubly degenerate, and energy bands with positive parity are isolated according to the character table of $C_{4v}$ PGS. For conventions, the four energy bands in Fig.~2(a) can be named as $s$, $p_x$, $p_y$ and $d_{xy}$, which are arranged from bottom to top accordingly. It is noticed that in Fig.~2(a) degenerate $p$ and $s$, $d_{xy}$ are separated by pseudo local gaps \cite{notation}.
\begin{figure}[h]
\begin{center}
\leavevmode
\includegraphics[clip=true,width=0.95\columnwidth]{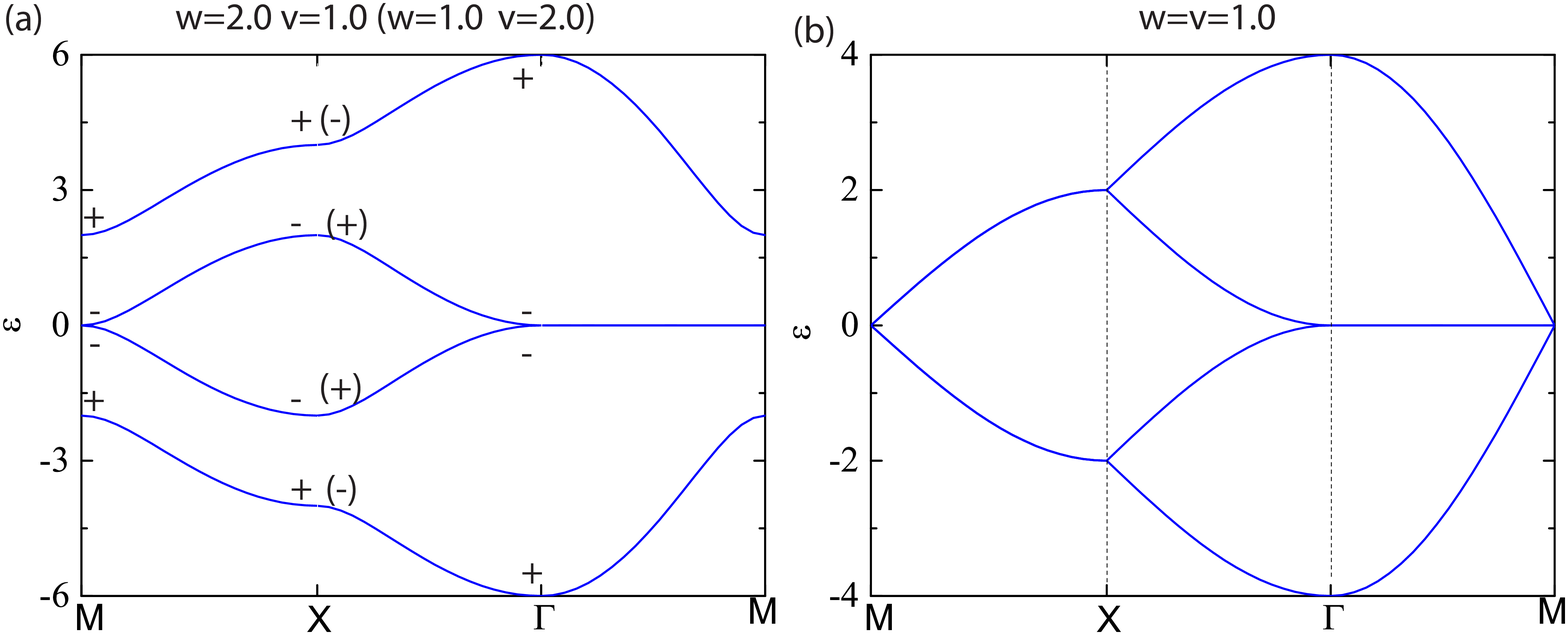}
\caption{Bulk spectra of 2D SSH model. From the lowest to the highest the energy bands are $s$, $p_x$, $p_y$, and $d_{xy}$, accordingly. The two $p$ bands are degenerate at $C_{4v}$ invariant points. Topological phase transition (band inversion) happens when intracellular hopping equals intercellular one. Bulk spectra before and after the band inversion is shown in (a), where the parities at high symmetric points before and after the band inversion are marked by "$\pm$" with and without parenthesis, respectively. The four bands touch at $X$ ($Y$) and $\Gamma$ points when $w=v=1.0$, which is shown in (b).}
\end{center}
\end{figure}

A topological phase transition associated with the 2D Zak phase happens at $|w|=|v|$. Here we show the energy spectrum at the critical point $w=v=1.0$ in Fig.~2(b). At this critical point, the four energy bands touch at $\Gamma$ and $X$ ($Y$) points. The band structure after the band inversion for $w=1.0$ and $v=2.0$ is same as Fig.~2(a) due to the sublattice symmetry. The topology of energy bands is encoded by their parities at high symmetric points, which are marked as "$\pm$" in Fig.~2(a). As displayed in Fig.~2(a), the parity of the lowest $s$ band at X (Y) point changes its sign after the band inversion, which indicates a topological phase transition. This topological phase transition can be characterized by the extended Zak phase in 2D, which is also the wave polarization given by the following expression \cite{Resta1994}:
\begin{equation}
\mathbf{P}=\frac{1}{2\pi}\int dk_x dk_y \text{Tr}[\mathbf{A}(k_x,k_y)],
\end{equation}
where $\mathbf{A}=\left<\psi|i\partial_\mathbf{k}|\psi\right>$ is Berry connection, and the integration is over the first BZ. Inversion symmetry put a strong constraint on the value of $\mathbf{P}$, which is determined gauge-independently by the parities at $\Gamma$ and $X$ ($Y$) points as \cite{Fang2012a}
\begin{equation}
P_i=\dfrac{1}{2}\left(\sum_n q_i^n \text{   }\text{modulo} \text{  } 2\right), \quad (-1)^{q^n_i}=\dfrac{\eta(X_i)}{\eta(\Gamma)},
\end{equation}
where $\eta$ denotes the parity, and the summation is over all the occupied bands, and $i$ stands for $x$ or $y$. Because of $C_4$ symmetry, we have $P_x=P_y$. After the band inversion one obtains $\mathbf{P}=(1/2,1/2)$ for the two topological nontrivial band gaps based on Eq.~(4).

Different from other TIs \cite{Haldane1988,Kane2005, Bernevig2006,Fu2011}, Berry curvature $\mathcal{F}(\mathbf{k})=\nabla \times \mathbf{A}$ of Eq.~(1)  vanishes everywhere in BZ due to the coexistence of TR and inversion symmetries. Because TR symmetry requires $\mathcal{F(-\mathbf{k})}=-\mathcal{F(\mathbf{k})}$ while the inversion symmetry require $\mathcal{F(-\mathbf{k})}=\mathcal{F(\mathbf{k})}$, which yields $-\mathcal{F}=\mathcal{F}=0$\cite{Xiao2010}. For above argument, the degeneracy of two p bands is not taken into consideration, and they induce singularities in Berry curvature even with simultaneous presences of TR and inversion symmetries. This energy degeneracy induced singularities of Berry curvature \emph{oscillate fast} along the $C_{4v}$-invariant line $|k_x|=|k_y|$ \cite{Supplement}, which gives a total zero integration over the first Brillouin zone. These singularities plays a similar role as magnetic solenoid in the AB effect or electronic degeneracy in the MAB effect \cite{Mead1992}. Besides the vanishing of Berry curvature of non-degenerate wavefunction, one cannot construct two pairs of pseudospins by the combinations of these four orbitals like the topological crystalline insulators \cite{Fu2011} as $s$ and $d_{xy}$ bands are not degenerate. Also, the conventional $Z_2$ number of TRI TIs \cite{Kane2005B} given by the product of parities at TRI points  are always trivial here due to the degeneracy of $p$ modes at $\Gamma$ and $M$ points. Thus, the system is different from previous 2D TIs, and its topological phase is characterized by the 2D Zak phase $(\pi, \pi)$ accompanying with the fractional wave polarization $(1/2,1/2)$. We want to emphasize that the 2D Zak phase or the wave polarization here are totally determined by the bulk property, different from the edge geometry effect studied in ref.\onlinecite{Delplace2011}, where the 1D Zak phase is determined by the edge shapes.

One essential feature of TIs is the existence of topological edge states when the system is in non-trivial topological phases. This feature is also manifested in our system. Edge states appear naturally when the $|w|<|v|$, for example $w=0$ as shown in Fig.~3(a). For clarity, we color the four atoms in the same unit cell in Fig.~3(a). It is noticed that there are four types of atoms in the edges, thus there should be totally four edge states. Figures 3(b) and 3(c) display the ribbon spectra of trivial and non-trivial phases, respectively. Because $s$ and $d_{xy}$ bands are not degenerate, at least two topological gaps are opening simultaneously in our $C_{4v}$ respected system. The edge states are marked as blue lines in Fig.~3(c). It is noticed that the edge states have quadratic dispersions here due to the $C_4$ PGS, and the ribbon spectra is gapless due to the lacking of completes band gaps in bulk spectrum as displayed in Fig.~1 \cite{Supplement2}. From Fig.~3(c) we observe two edge states from the ribbon spectrum in total. However we have four distinct bands in our system, and they are all involved in the topological phase transition. Thus the edge states are doubly degenerate. The wavefunctions of doubly degenerate edge states are plotted in Fig.~3(d), where one sees that the degenerate edge states are symmetric and anti-symmetric along the $x$ direction. This degeneracy is caused by the inversion symmetry along the $x$ direction. The inset of Fig.~3(d) displays the spatial distribution of the degenerate edge states, which coincides with each other. The degeneracy of edge states in this 2D system may yield non-Abellian statics \cite{Frohlich1988}, which remains as a future study topic. 
\begin{figure}[h]
\begin{center}
\leavevmode
\includegraphics[clip=true,width=0.95\columnwidth]{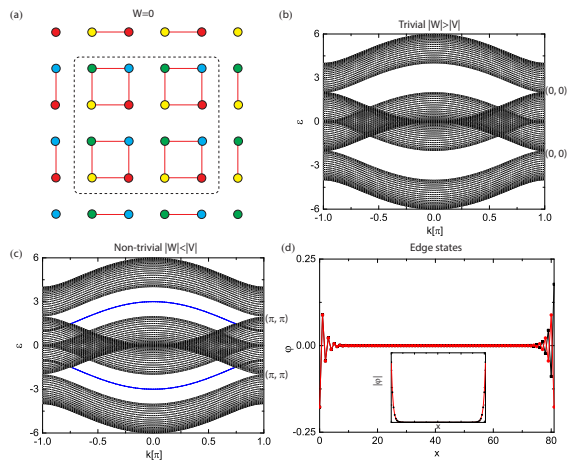}
\caption{(a) Edge states naturally appear when the intracellular hopping is 0. There are four types of of atoms (marked as different colors in a unit cell) in the edge, thus there should be four edge states in total. Ribbon spectra for (b) $w=2.0$, $v=1.0$ and (c) $w=1.0$, $v=2.0$, respectively. After the band inversion, topological edge states marked as blue lines appear in the two topological gaps, which are characterized by the two dimensional Zak phase $(\pi, \pi)$ accompanying the fractional wave polarization $(1/2, 1/2)$ for each topological gap as marked by the right side of the panel. The topologically trivial gaps have 2D Zak phase $(0, 0)$. Wavefunction of edge states at $k=0.2\pi$ is shown in (d). The edge states are doubly degenerate due to the inversion symmetry along $x$ direction, which are symmetric and anti-symmetric along the $x$ direction. The spatial distribution of the degenerate edge states are shown in inset of (d).}
\end{center}
\end{figure}

Let us briefly discuss the robustness of the topological edge states. As investigated in the supplement \cite{Supplement3}, the difference between two hoppings $|v|-|w|$ determines the robustness of the edge states. For perturbations such as surface potentials and nonuniform on-site energies with amplitudes smaller than $|v|-|w|$, the edge states always exist in the band gaps even for those possessing local band gaps, which is different from the case discussed in ref.\onlinecite{Miert2016}. Furthermore, as the $(\pi, \pi)$ phase is protected by the $C_{4v}$ PGS, the edge states can appear in a general direction like $x=y$ direction. It should be noticed that the band structure of edge states strongly depend on the directions of the ribbons. One may also concern that the spectrum of edge states shown do not bridge valence and conduction bands, which makes the edge states seemingly trivial. This is because of open (vacuum) boundary condition used in the calculation, which breaks the $C_{4v}$ PGS. For a nontrivial ribbon bounded by another trivial ribbon possessing $C_{4v}$ symmetry, the edge states bridge the valence and conduction bands.

Finally, we plot the topological phase diagram of our model in Fig.~4. When the intercellular hopping is larger than the intracellular one; the system enters the non-trivial topological phase. The nontrivial/trivial topological phases are characterized by the 2D Zak phase $(\pi,\pi)$/$(0, 0)$ accompanying with fractional wave polarization $(1/2,1/2)$/$(0, 0)$.
\begin{figure}[h]
\begin{center}
\leavevmode
\includegraphics[clip=true,width=0.6\columnwidth]{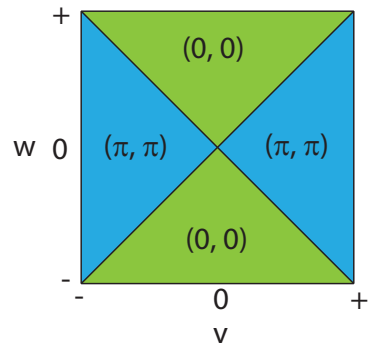}
\caption{Topological phase diagram of the 2D SSH model. The system enters topological nontrivial phase when $|w|<|v|$. The nontrivial topological phases are characterized by the nonzero 2D Zak phase $(\pi, \pi)$. }
\end{center}
\end{figure}

Although we take the $C_{4v}$ PGS as an example here, the results can be extended to other PGS straightforwardly. For example, the $C_{2v}$ PGS, in which $P_x=P_y$ is not a necessary requirement, and the appearance of edge states will depend on the cutting direction of bulk samples. Besides the square lattice, another case is the hexagonal lattice. For hexagonal case or $C_{6v}$ PGS, there are two pairs of degenerate bands with opposite parities, and a pseudospin-polarized state can be constructed \cite{Wu2015}. The topological phases of the hexagonal case can be described by a pseudospin Chern number then. But no matter what kind of lattice is under consideration, the topological phase transition always happens at $|w|=|v|$. It is noticed that we consider electrons here, but the results can be extended to another system such as dielectric photonic crystals \cite{Liu2016}, which can be more easily to be realized by experimentalists due to their macroscopic system sizes.

To summarize, we have shown nontrivial topological phases in the 2D lattice system with zero Berry curvature, which is descried by the SSH model in two dimensions. The non-trivial topological phase of 2D SSH model have been characterized by the 2D Zak phase accompanying with the fractional wave polarizations. Bulk edge correspondence associated with the 2D Zak phase is manifested. Our results will serve to design the new materials with a nontrivial topological phase.

F. Liu appreciate the beneficial discussions with H.-Y. Deng, Y. Xu, Zh. Li, M. Lein, and F. Nori. We also thank all the referees for the useful discussions and their insightful suggestions. This work is supported by JSPS KAKENHI Grant No. 15K13507 and MEXT KAKENHI Grant No. 25107005.


\begin{thebibliography}{21}%
\makeatletter
\providecommand \@ifxundefined [1]{%
 \@ifx{#1\undefined}
}%
\providecommand \@ifnum [1]{%
 \ifnum #1\expandafter \@firstoftwo
 \else \expandafter \@secondoftwo
 \fi
}%
\providecommand \@ifx [1]{%
 \ifx #1\expandafter \@firstoftwo
 \else \expandafter \@secondoftwo
 \fi
}%
\providecommand \natexlab [1]{#1}%
\providecommand \enquote  [1]{``#1''}%
\providecommand \bibnamefont  [1]{#1}%
\providecommand \bibfnamefont [1]{#1}%
\providecommand \citenamefont [1]{#1}%
\providecommand \href@noop [0]{\@secondoftwo}%
\providecommand \href [0]{\begingroup \@sanitize@url \@href}%
\providecommand \@href[1]{\@@startlink{#1}\@@href}%
\providecommand \@@href[1]{\endgroup#1\@@endlink}%
\providecommand \@sanitize@url [0]{\catcode `\\12\catcode `\$12\catcode
  `\&12\catcode `\#12\catcode `\^12\catcode `\_12\catcode `\%12\relax}%
\providecommand \@@startlink[1]{}%
\providecommand \@@endlink[0]{}%
\providecommand \url  [0]{\begingroup\@sanitize@url \@url }%
\providecommand \@url [1]{\endgroup\@href {#1}{\urlprefix }}%
\providecommand \urlprefix  [0]{URL }%
\providecommand \Eprint [0]{\href }%
\providecommand \doibase [0]{http://dx.doi.org/}%
\providecommand \selectlanguage [0]{\@gobble}%
\providecommand \bibinfo  [0]{\@secondoftwo}%
\providecommand \bibfield  [0]{\@secondoftwo}%
\providecommand \translation [1]{[#1]}%
\providecommand \BibitemOpen [0]{}%
\providecommand \bibitemStop [0]{}%
\providecommand \bibitemNoStop [0]{.\EOS\space}%
\providecommand \EOS [0]{\spacefactor3000\relax}%
\providecommand \BibitemShut  [1]{\csname bibitem#1\endcsname}%
\let\auto@bib@innerbib\@empty
%</preamble>
\bibitem [{\citenamefont {Hasan}\ and\ \citenamefont {Kane}(2010)}]{Hasan2010}%
  \BibitemOpen
  \bibfield  {author} {\bibinfo {author} {\bibfnamefont {M.~Z.}\ \bibnamefont
  {Hasan}}\ and\ \bibinfo {author} {\bibfnamefont {C.~L.}\ \bibnamefont
  {Kane}},\ }\href {\doibase 10.1103/RevModPhys.82.3045} {\bibfield  {journal}
  {\bibinfo  {journal} {Rev. Mod. Phys.}\ }\textbf {\bibinfo {volume} {82}},\
  \bibinfo {pages} {3045} (\bibinfo {year} {2010})}\BibitemShut {NoStop}%
\bibitem [{\citenamefont {Qi}\ and\ \citenamefont {Zhang}(2011)}]{Qi2011}%
  \BibitemOpen
  \bibfield  {author} {\bibinfo {author} {\bibfnamefont {X.-L.}\ \bibnamefont
  {Qi}}\ and\ \bibinfo {author} {\bibfnamefont {S.-C.}\ \bibnamefont {Zhang}},\
  }\href {\doibase 10.1103/RevModPhys.83.1057} {\bibfield  {journal} {\bibinfo
  {journal} {Rev. Mod. Phys.}\ }\textbf {\bibinfo {volume} {83}},\ \bibinfo
  {pages} {1057} (\bibinfo {year} {2011})}\BibitemShut {NoStop}%
\bibitem [{\citenamefont {Bansil}\ \emph {et~al.}(2016)\citenamefont {Bansil},
  \citenamefont {Lin},\ and\ \citenamefont {Das}}]{Bansil2016}%
  \BibitemOpen
  \bibfield  {author} {\bibinfo {author} {\bibfnamefont {A.}~\bibnamefont
  {Bansil}}, \bibinfo {author} {\bibfnamefont {H.}~\bibnamefont {Lin}}, \ and\
  \bibinfo {author} {\bibfnamefont {T.}~\bibnamefont {Das}},\ }\href {\doibase
  10.1103/RevModPhys.88.021004} {\bibfield  {journal} {\bibinfo  {journal}
  {Rev. Mod. Phys.}\ }\textbf {\bibinfo {volume} {88}},\ \bibinfo {pages}
  {021004} (\bibinfo {year} {2016})}\BibitemShut {NoStop}%
\bibitem [{\citenamefont {Klitzing}\ \emph {et~al.}(1980)\citenamefont
  {Klitzing}, \citenamefont {Dorda},\ and\ \citenamefont
  {Pepper}}]{Klitzing1980}%
  \BibitemOpen
  \bibfield  {author} {\bibinfo {author} {\bibfnamefont {K.~v.}\ \bibnamefont
  {Klitzing}}, \bibinfo {author} {\bibfnamefont {G.}~\bibnamefont {Dorda}}, \
  and\ \bibinfo {author} {\bibfnamefont {M.}~\bibnamefont {Pepper}},\ }\href
  {\doibase 10.1103/PhysRevLett.45.494} {\bibfield  {journal} {\bibinfo
  {journal} {Phys. Rev. Lett.}\ }\textbf {\bibinfo {volume} {45}},\ \bibinfo
  {pages} {494} (\bibinfo {year} {1980})}\BibitemShut {NoStop}%
\bibitem [{\citenamefont {Haldane}(1988)}]{Haldane1988}%
  \BibitemOpen
  \bibfield  {author} {\bibinfo {author} {\bibfnamefont {F.~D.~M.}\
  \bibnamefont {Haldane}},\ }\href {\doibase 10.1103/PhysRevLett.61.2015}
  {\bibfield  {journal} {\bibinfo  {journal} {Phys. Rev. Lett.}\ }\textbf
  {\bibinfo {volume} {61}},\ \bibinfo {pages} {2015} (\bibinfo {year}
  {1988})}\BibitemShut {NoStop}%
\bibitem [{\citenamefont {Kane}\ and\ \citenamefont
  {Mele}(2005{\natexlab{a}})}]{Kane2005}%
  \BibitemOpen
  \bibfield  {author} {\bibinfo {author} {\bibfnamefont {C.~L.}\ \bibnamefont
  {Kane}}\ and\ \bibinfo {author} {\bibfnamefont {E.~J.}\ \bibnamefont
  {Mele}},\ }\href {\doibase 10.1103/PhysRevLett.95.226801} {\bibfield
  {journal} {\bibinfo  {journal} {Phys. Rev. Lett.}\ }\textbf {\bibinfo
  {volume} {95}},\ \bibinfo {pages} {226801} (\bibinfo {year}
  {2005}{\natexlab{a}})}\BibitemShut {NoStop}%
\bibitem [{\citenamefont {Bernevig}\ \emph {et~al.}(2006)\citenamefont
  {Bernevig}, \citenamefont {Hughes},\ and\ \citenamefont
  {Zhang}}]{Bernevig2006}%
  \BibitemOpen
  \bibfield  {author} {\bibinfo {author} {\bibfnamefont {B.~A.}\ \bibnamefont
  {Bernevig}}, \bibinfo {author} {\bibfnamefont {T.~L.}\ \bibnamefont
  {Hughes}}, \ and\ \bibinfo {author} {\bibfnamefont {S.-C.}\ \bibnamefont
  {Zhang}},\ }\href {\doibase 10.1126/science.1133734} {\bibfield  {journal}
  {\bibinfo  {journal} {Science}\ }\textbf {\bibinfo {volume} {314}},\ \bibinfo
  {pages} {1757} (\bibinfo {year} {2006})}\BibitemShut {NoStop}%
\bibitem [{\citenamefont {Fu}(2011)}]{Fu2011}%
  \BibitemOpen
  \bibfield  {author} {\bibinfo {author} {\bibfnamefont {L.}~\bibnamefont
  {Fu}},\ }\href {\doibase 10.1103/PhysRevLett.106.106802} {\bibfield
  {journal} {\bibinfo  {journal} {Phys. Rev. Lett.}\ }\textbf {\bibinfo
  {volume} {106}},\ \bibinfo {pages} {106802} (\bibinfo {year}
  {2011})}\BibitemShut {NoStop}%
\bibitem [{\citenamefont {Qi}\ \emph {et~al.}(2008)\citenamefont {Qi},
  \citenamefont {Hughes},\ and\ \citenamefont {Zhang}}]{Qi2008}%
  \BibitemOpen
  \bibfield  {author} {\bibinfo {author} {\bibfnamefont {X.-L.}\ \bibnamefont
  {Qi}}, \bibinfo {author} {\bibfnamefont {T.~L.}\ \bibnamefont {Hughes}}, \
  and\ \bibinfo {author} {\bibfnamefont {S.-C.}\ \bibnamefont {Zhang}},\ }\href
  {\doibase 10.1103/PhysRevB.78.195424} {\bibfield  {journal} {\bibinfo
  {journal} {Phys. Rev. B}\ }\textbf {\bibinfo {volume} {78}},\ \bibinfo
  {pages} {195424} (\bibinfo {year} {2008})}\BibitemShut {NoStop}%
\bibitem [{\citenamefont {Sheng}\ \emph {et~al.}(2006)\citenamefont {Sheng},
  \citenamefont {Weng}, \citenamefont {Sheng},\ and\ \citenamefont
  {Haldane}}]{Haldane2006}%
  \BibitemOpen
  \bibfield  {author} {\bibinfo {author} {\bibfnamefont {D.~N.}\ \bibnamefont
  {Sheng}}, \bibinfo {author} {\bibfnamefont {Z.~Y.}\ \bibnamefont {Weng}},
  \bibinfo {author} {\bibfnamefont {L.}~\bibnamefont {Sheng}}, \ and\ \bibinfo
  {author} {\bibfnamefont {F.~D.~M.}\ \bibnamefont {Haldane}},\ }\href
  {\doibase 10.1103/PhysRevLett.97.036808} {\bibfield  {journal} {\bibinfo
  {journal} {Phys. Rev. Lett.}\ }\textbf {\bibinfo {volume} {97}},\ \bibinfo
  {pages} {036808} (\bibinfo {year} {2006})}\BibitemShut {NoStop}%
\bibitem [{\citenamefont {Shiozaki}\ and\ \citenamefont
  {Sato}(2014)}]{Shiozaki2014}%
  \BibitemOpen
  \bibfield  {author} {\bibinfo {author} {\bibfnamefont {K.}~\bibnamefont
  {Shiozaki}}\ and\ \bibinfo {author} {\bibfnamefont {M.}~\bibnamefont
  {Sato}},\ }\href {\doibase 10.1103/PhysRevB.90.165114} {\bibfield  {journal}
  {\bibinfo  {journal} {Phys. Rev. B}\ }\textbf {\bibinfo {volume} {90}},\
  \bibinfo {pages} {165114} (\bibinfo {year} {2014})}\BibitemShut {NoStop}%
\bibitem [{\citenamefont {Dong}\ and\ \citenamefont {Liu}(2016)}]{Dong2016}%
  \BibitemOpen
  \bibfield  {author} {\bibinfo {author} {\bibfnamefont {X.-Y.}\ \bibnamefont
  {Dong}}\ and\ \bibinfo {author} {\bibfnamefont {C.-X.}\ \bibnamefont {Liu}},\
  }\href {\doibase 10.1103/PhysRevB.93.045429} {\bibfield  {journal} {\bibinfo
  {journal} {Phys. Rev. B}\ }\textbf {\bibinfo {volume} {93}},\ \bibinfo
  {pages} {045429} (\bibinfo {year} {2016})}\BibitemShut {NoStop}%
\bibitem [{\citenamefont {Xiao}\ \emph {et~al.}(2010)\citenamefont {Xiao},
  \citenamefont {Chang},\ and\ \citenamefont {Niu}}]{Xiao2010}%
  \BibitemOpen
  \bibfield  {author} {\bibinfo {author} {\bibfnamefont {D.}~\bibnamefont
  {Xiao}}, \bibinfo {author} {\bibfnamefont {M.-C.}\ \bibnamefont {Chang}}, \
  and\ \bibinfo {author} {\bibfnamefont {Q.}~\bibnamefont {Niu}},\ }\href
  {\doibase 10.1103/RevModPhys.82.1959} {\bibfield  {journal} {\bibinfo
  {journal} {Rev. Mod. Phys.}\ }\textbf {\bibinfo {volume} {82}},\ \bibinfo
  {pages} {1959} (\bibinfo {year} {2010})}\BibitemShut {NoStop}%
\bibitem [{\citenamefont {Thouless}(2011)}]{book2}%
  \BibitemOpen
  \bibfield  {author} {\bibinfo {author} {\bibfnamefont {D.~J.}\ \bibnamefont
  {Thouless}},\ }\enquote {\bibinfo {title} {Quantization of electric
  charge},}\ in\ \href@noop {} {\emph {\bibinfo {booktitle} {Topological
  Quantum Numbers in Nonrelativistic Physics}}}\ (\bibinfo  {publisher} {WORLD
  SCIENTIFIC},\ \bibinfo {year} {2011})\ Chap.~\bibinfo {chapter} {2}, pp.\
  \bibinfo {pages} {16--20}\BibitemShut {NoStop}%
\bibitem{Mead1979}
{C. A. Mead, Chem. Phys. \textbf{49}, 23 (1980).}
\bibitem{Mead1992}
{C. A. Mead, Rev. Mod. Phys. \textbf{64}, 51 (1992).}

\bibitem{Hughes2011}
{T. L. Hughes, E. Prodan, and B. A. Bernevig, Phys. Rev. B \text{83}, 245132 (2011)}

\bibitem{Lau2015}
{A. Lau, C. Ortix, and J. van den Brink, Phys. Rev. Lett. \text{115}, 216805 (2015)}
\bibitem{Lau2016}
{A. Lau, J. van den Brink, and C. Ortix, Phys. Rev. B \text{94}, 165164 (2016)}

\bibitem [{\citenamefont {Zak}(1989)}]{Zak1989}%
  \BibitemOpen
  \bibfield  {author} {\bibinfo {author} {\bibfnamefont {J.}~\bibnamefont
  {Zak}},\ }\href {\doibase 10.1103/PhysRevLett.62.2747} {\bibfield  {journal}
  {\bibinfo  {journal} {Phys. Rev. Lett.}\ }\textbf {\bibinfo {volume} {62}},\
  \bibinfo {pages} {2747} (\bibinfo {year} {1989})}\BibitemShut {NoStop}%
\bibitem{notation}
{$s$, $d_{xy}$ bands can also be real (complete) gapped for other parameters like $w=1.0$ and $v=3.0$.}
\bibitem [{\citenamefont {Resta}(1994)}]{Resta1994}%
  \BibitemOpen
  \bibfield  {author} {\bibinfo {author} {\bibfnamefont {R.}~\bibnamefont
  {Resta}},\ }\href {\doibase 10.1103/RevModPhys.66.899} {\bibfield  {journal}
  {\bibinfo  {journal} {Rev. Mod. Phys.}\ }\textbf {\bibinfo {volume} {66}},\
  \bibinfo {pages} {899} (\bibinfo {year} {1994})}\BibitemShut {NoStop}%
\bibitem [{\citenamefont {Fang}\ \emph {et~al.}(2012)\citenamefont {Fang},
  \citenamefont {Gilbert},\ and\ \citenamefont {Bernevig}}]{Fang2012a}%
  \BibitemOpen
  \bibfield  {author} {\bibinfo {author} {\bibfnamefont {C.}~\bibnamefont
  {Fang}}, \bibinfo {author} {\bibfnamefont {M.~J.}\ \bibnamefont {Gilbert}}, \
  and\ \bibinfo {author} {\bibfnamefont {B.~A.}\ \bibnamefont {Bernevig}},\
  }\href {\doibase 10.1103/PhysRevB.86.115112} {\bibfield  {journal} {\bibinfo
  {journal} {Phys. Rev. B}\ }\textbf {\bibinfo {volume} {86}},\ \bibinfo
  {pages} {115112} (\bibinfo {year} {2012})}\BibitemShut {NoStop}%
\bibitem [{\citenamefont {Kane}\ and\ \citenamefont
  {Mele}(2005{\natexlab{b}})}]{Kane2005B}%
  \BibitemOpen
  \bibfield  {author} {\bibinfo {author} {\bibfnamefont {C.~L.}\ \bibnamefont
  {Kane}}\ and\ \bibinfo {author} {\bibfnamefont {E.~J.}\ \bibnamefont
  {Mele}},\ }\href {\doibase 10.1103/PhysRevLett.95.146802} {\bibfield
  {journal} {\bibinfo  {journal} {Phys. Rev. Lett.}\ }\textbf {\bibinfo
  {volume} {95}},\ \bibinfo {pages} {146802} (\bibinfo {year}
  {2005}{\natexlab{b}})}\BibitemShut {NoStop}%
\bibitem{Supplement}
{The fast oscillation of Berry curavture in the momentum space gives zero Chern number, coindciding with the TR symmetry and inversion symmetry. See the Berry curvature of p band in the supplement.}
\bibitem{Delplace2011}
{P. Delplace, D. Ullmo, and G. Montambaux, Phys. Rev. B \textbf{84}, 195452 (2011).}
\bibitem{Supplement2}
{For ribbon spectra of nontrivial topological phase having large real band gaps and extra small local band gaps at other parameters, please refer to the supplement.}

\bibitem [{\citenamefont {Fr{\"o}hlich}(1988)}]{Frohlich1988}%
  \BibitemOpen
  \bibfield  {author} {\bibinfo {author} {\bibfnamefont {J.}~\bibnamefont
  {Fr{\"o}hlich}},\ }\enquote {\bibinfo {title} {Statistics of fields, the
  yang-baxter equation, and the theory of knots and links},}\ in\ \href
  {\doibase 10.1007/978-1-4613-0729-7_4} {\emph {\bibinfo {booktitle}
  {Nonperturbative Quantum Field Theory}}},\ \bibinfo {editor} {edited by\
  \bibinfo {editor} {\bibfnamefont {G.}~\bibnamefont {'t~Hooft}}, \bibinfo
  {editor} {\bibfnamefont {A.}~\bibnamefont {Jaffe}}, \bibinfo {editor}
  {\bibfnamefont {G.}~\bibnamefont {Mack}}, \bibinfo {editor} {\bibfnamefont
  {P.~K.}\ \bibnamefont {Mitter}}, \ and\ \bibinfo {editor} {\bibfnamefont
  {R.}~\bibnamefont {Stora}}}\ (\bibinfo  {publisher} {Springer US},\ \bibinfo
  {address} {Boston, MA},\ \bibinfo {year} {1988})\ pp.\ \bibinfo {pages}
  {71--100}\BibitemShut {NoStop}%
\bibitem{Supplement3}
{For ribbon spectra of nontrivial topological phase in presences of various perturbations such as surface potentials and nonuniform on-site energies, ribbon spectrum in $x=y$ direction, and ribbon spectrum of a nontrivial sample bounded by a trivial one possessing $C_{4v}$ symmetry, please refer to the supplement.}
\bibitem{Miert2016}
{G. van Miert, C. Ortix, C. M. Smith, 2D Materials \text{4}, 015023 (2016)}
\bibitem [{\citenamefont {Wu}\ and\ \citenamefont {Hu}(2015)}]{Wu2015}%
  \BibitemOpen
  \bibfield  {author} {\bibinfo {author} {\bibfnamefont {L.-H.}\ \bibnamefont
  {Wu}}\ and\ \bibinfo {author} {\bibfnamefont {X.}~\bibnamefont {Hu}},\ }\href
  {\doibase 10.1103/PhysRevLett.114.223901} {\bibfield  {journal} {\bibinfo
  {journal} {Phys. Rev. Lett.}\ }\textbf {\bibinfo {volume} {114}},\ \bibinfo
  {pages} {223901} (\bibinfo {year} {2015})}\BibitemShut {NoStop}%
\bibitem [{\citenamefont {Liu}\ \emph {et~al.}()\citenamefont {Liu} \emph
  {et~al.}}]{Liu2016}%
  \BibitemOpen
  \bibfield  {author} {\bibinfo {author} {\bibfnamefont {F.}~\bibnamefont
  {Liu}} \emph {et~al.},\ }\href@noop {} {\bibinfo  {journal} {unpublished}\
  }\BibitemShut {NoStop}%
\end{thebibliography}
\end{document}